\documentclass[reprint,amsmath,amssymb,aps,superscriptaddress]{revtex4-2}
\usepackage{graphicx}
\usepackage{dcolumn}
\usepackage{bm}

\begin{document}

\title{Modulation instability of Kerr optical frequency combs in dual-coupled optical cavities}

\author{Enxu Zhu}
\affiliation{%
College of Science, Hangzhou Dianzi University, Zhejiang 310018, China
}%

\author{Chaoying Zhao}%
\email{Corresponding author: zchy49@163.com}
\affiliation{%
College of Science, Hangzhou Dianzi University, Zhejiang 310018, China
}%
\affiliation{%
State Key Laboratory of Quantum Optics and Quantum Optics Devices, Institute of Opto-Electronics, Shanxi University, Taiyuan 030006, China
}%

\date{\today}

\begin{abstract}
Kerr optical frequency combs generated in a coherently driven Kerr nonlinear resonator has the potential for a wide range of applications. However, in a single cavity which is a widely adopted configuration for Kerr optical frequency combs generation, modulation instability is suppressed in normal dispersion regime and the pump-to-comb conversion efficiency is extremely low for a single dissipative Kerr soliton in anomalous dispersion regime. Dual-coupled cavities have been proposed to generate Kerr optical frequency combs in normal dispersion regime, and have potential to remarkably increase conversion efficiency for Kerr optical frequency combs. Here, we investigate modulation instability and Kerr optical frequency-comb formation in dual-coupled cavities. Based on solutions of the continuous-wave steady state, we obtain a quadric algebraic equation describing the modulation instability gain, and we find that it is intensely influenced by the group velocity mismatch between the two cavities. Our numerical simulations demonstrate that platicons can be generated via pump scanning scheme for the case that both the two cavities possess normal dispersion, and a single dissipative Kerr soliton can be generated in the cavity with anomalous dispersion while the dispersion of the other cavity is normal. Our analysis of modulation instability provides a powerful tool for Kerr optical frequency-comb generation via pump modulation and cavity detuning tuning scheme in dual-coupled cavities.
\end{abstract}

\maketitle

\section{\label{sec:level1}Introduction}
Kerr optical frequency combs (OFCs) generated by pumped Kerr nonlinear optical cavities \cite{Haye2007, Ferdous2011, Stern2018, Xiang2021} have attracted significant interest for more than a decade due to their high degree compactness. Mode-locked Kerr OFCs usually attach to periodic temporal patterns or, especially, dissipative Kerr solitons (DKSs) in anomalous dispersion regime \cite{Kippenberg2018}. As the special solutions of the nonlinear Schr\"{o}dinger equation (NLSE), solitons have been widely studied in different contexts such as Bose-Einstein condensates (BEC) \cite{Kengne2021}, optical communications \cite{Haus1996}, and mode-locked lasers \cite{Grelu2012}. Since Kerr OFCs associating with DKSs were first observed in a ${\rm MgF_2}$ crystal microresonator in 2014 \cite{Herr2014}, DKSs have been demonstrated in several platforms including $\rm{Si_3N_4}$ \cite{Ye2019}, $\rm{AlN}$ \cite{Gong2018}, and silica \cite{Yi2015}. Meanwhile, the influence of high order dispersion \cite{Cherenkov2017, Bao2017, Parra2017}, dispersion perturbation caused by mode coupling between different mode families \cite{Herr2014a, Jang2016}, two-component pump \cite{Bao2019, Zhang2020, Xu2021}, thermo-optic effect \cite{Guo2017, Valery2021}, and Raman effect \cite{Yang2017, Gong2020} on Kerr OFCs was investigated. With the development of silicon photonics \cite{Liu2021}, there are increasingly applications having been proposed based on on-chip Kerr OFCs, such as ultrafast distance measurements \cite{Trocha2018}, parallel convolution processing \cite{Feldmann2021}, and spectroscopy \cite{Dutt2018}.

The widely adopted configuration for Kerr OFCs is a single cavity coupled with a bus waveguide \cite{Levy2010, Herr2012, Brasch2016, Gaeta2019, Wang2020, Voloshin2021}. Despite the great advance in Kerr OFCs, in a single cavity system the pump-to-comb conversion efficiency of a single DKS is usually lower than a few percent \cite{Bao2014, Wang2016, Xue2017} due to the large detuning required for DKSs. Using dark pulses supported in normal dispersion regime indeed enhances the comb power \cite{Xue2017}. However, it has been demonstrated that normal dispersion is not conducive to the formation of Kerr OFCs in a single cavity system \cite{Chembo2010, Hansson2013}. The formation of a reported mode-locked dark pulse \cite{Xue2015} is induced by the local anomalous dispersion caused by the coupling between different mode families although the overall dispersion is normal. An alternative approach for enhancing the conversion efficiency is the use of a dual-coupled cavities system \cite{Xue2019}. In such a system, the pump power is almost entirely converted to the comb power assuming the two cavities are intrinsically lossless. Moreover, due to the coupling between the two cavities, it is possible for the generation of Kerr OFCs in normal dispersion regime \cite{Xue2015a}.

Modulation instability (MI), which breaks the continuous-wave (cw) steady state and stimulates the growth of sidebands, is the essential initial stage for OFC formation, and has been well studied in single cavity systems \cite{Hansson2013, Godey2014}. Coherently pumped single cavity systems are described by the so-called Lugiato-Lefever equation (LLE) \cite{Coen2013}, which is a mean-field approximation equation derived from Ikeda map \cite{Ikeda1979} that is the combination of NLSE and the cavity boundary condition. The MI analysis reveals that anomalous dispersion is critical for the generation of Kerr OFCs in a single cavity system \cite{Chembo2010, Hansson2013, Godey2014}. Besides, MI of orthogonal-polarization pumped single cavity systems has been theoretically \cite{Hansson2018} and experimentally \cite{Fatome2020} demonstrated. In addition to Kerr OFCs, MI of quadratic OFCs, which are generated by the quadratic nonlinearity (Kerr nonlinearity corresponds to the cubic nonlinearity), has been studied in various cases including second harmonic generation (SHG) combs \cite{Leo2016, Leo2016a, Hansson2017} and optical parametric oscillation (OPO) combs \cite{Mosca2018}.

Previous studies on Kerr OFCs generation in dual-coupled cavities systems are based on the analysis of mode coupling between the two cavities \cite{Xue2015a, Miller2015, Fujii2018}. However, the quantitative analysis of MI in dual-coupled cavities is still remained to be investigated. Our main aim in this work is to provide a new insight into MI for Kerr OFCs generation in dual-coupled cavities with arbitrary dispersion of each coupled cavity. In particular, we are most interested in two cases about the cavity dispersion. One of which is that both the two cavities possess normal dispersion, and the other is that one of the two cavities possesses anomalous dispersion while the dispersion of the other cavity is normal. On one hand, although many materials are used for on-chip Kerr OFCs, the dispersion of most materials is normal in the near-infrared band without elaborate dispersion engineering \cite{Okawachi2011, Riemensberger2012, Grudinin2015}. Using dual-coupled cavities systems would liberate the Kerr OFCs from its dependence on anomalous dispersion. On the other hand, Kerr OFCs associating with DKSs have a broad spectrum band and smooth spectrum envelope. As mentioned above, dual-coupled cavities configurations have been proposed to enhance the pump-to-comb conversion efficiency of DKSs \cite{Xue2019}, in which the cavity with normal dispersion termed as ``pump cavity" is used to recycle the pump power, and the cavity with anomalous dispersion termed as ``soliton cavity" is used to sustain DKSs. However, due to the lack of the knowledge of MI for dual-coupled cavities systems, the DKS in the previous work \cite{Xue2019} was excited by introducing a high peak power gaussian pulse, which needs a special amplification scheme such as chirped pulse amplification (CPA) \cite{Strickland1985} undoubtedly leading to increasing the complexity of dual-coupled cavities systems. This article is organized as follow. In Sec.~\ref{sec:level2}, we introduce the mean-field equations for Kerr OFCs in the dual-coupled cavities system. The cw steady state and MI of the system are analysed in Sec.~\ref{sec:level3}. Numerical simulation results with the beginning of MI in Sec.~\ref{sec:level4} present a platicon in normal dispersion regime, and a DKS in anomalous dispersion regime. Finally, the article is concluded in the last section.

\section{\label{sec:level2}Theoretical model}
\begin{figure}[tb]
\includegraphics{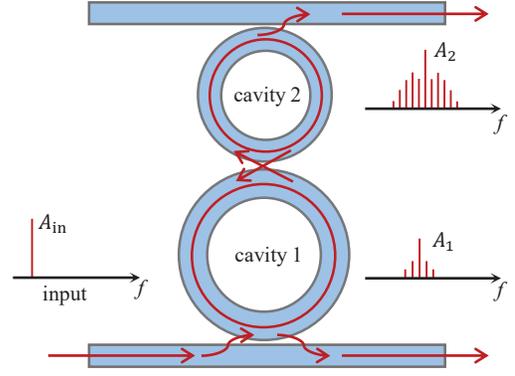}
\caption{\label{fig:1}Schematic of the OFC generation in dual-coupled cavities. The continuous-wave pump $A_{\rm in}$ is coupled into the cavity 1 through a bus waveguide. The Kerr nonlinearity of the system gives rise to optical frequency combs $A_1$ in the cavity 1 and $A_2$ in the cavity 2.}
\end{figure}
Fig.~\ref{fig:1} shows a typical dual-coupled cavities configuration for OFC generation. The external pump is coupled into the cavity 1 through the bus waveguide, and is delivered to the cavity 2 through the coupling between the two cavities. The complex field amplitudes $A_{1,2}$ in the cavity 1 and the cavity 2, respectively, satisfy the following coupled mean-field equations \cite{Xue2019}
\begin{eqnarray}
\frac{\partial A_{1}}{\partial z}=&&\left[-\alpha_{1}-i\delta_{1}-\Delta k^{'}\frac{\partial}{\partial \tau}-i\frac{k_1^{''}}{2}\frac{\partial^2}{\partial \tau^2}+i\gamma_{1}|A_{1}|^2\right]A_{1}\nonumber\\
&&+i\kappa_{\rm c}A_{2}+i\kappa_{1}A_{\rm in},
\label{eq:1}
\end{eqnarray}
\begin{equation}
\frac{\partial A_{2}}{\partial z}=\left[-\alpha_{2}-i\delta_{2}-i\frac{k_{2}^{''}}{2}\frac{\partial^2}{\partial \tau^2}+i\gamma_{2}|A_{2}|^2\right]A_{2}+i\kappa_{\rm c}A_{1},
\label{eq:2}
\end{equation}
where $z$ is the propagation distance in the cavities; $\tau$ is the so-called fast time \cite{Coen2013}; $A_{\rm in}$ is the pump field; $\delta_{1,2}$ are the averaged frequency detunings; $\alpha_{1,2}=\alpha_{\rm i1,2}-{\rm ln}(1-\theta_{\rm 1,2})/2L_2$ are the total cavity linear losses in the cavity 1 and the cavity 2, respectively, where $\alpha_{\rm i2}$ is the intrinsic propagation losses in the cavity 2, and $\theta_{1,2}$ are the power coupling ratio between the cavities and their corresponding bus waveguide, and $L_2$ are the length of the cavity 1. We note that parameters in Eq.~(\ref{eq:1}) are scaled according to the length ratio of the cavity 1 to the cavity 2, $r=L_1/L_2$ with $L_1$ the length of the cavity 1, so that $\alpha_{\rm i1}=r\alpha_{\rm i1}^{'}$ with $\alpha_{\rm i1}^{'}$ the intrinsic propagation losses in the cavity 1. $k_1^{''}=r d^2k_1/d\omega^2|_{\omega_0}$ and $k_2^{''}=d^2k_2/d\omega^2|_{\omega_0}$ are the second-order dispersion coefficients with $k_{1,2}$ the magnitude of the wave vector and $\omega_0$ the frequency of the pump; $\Delta k^{'}=r dk_{1}/d\omega|_{\omega_0}-dk_{2}/d\omega|_{\omega_0}$ is the group velocity mismatch; $\gamma_{1}=r\gamma_{1}^{'}$ with $\gamma_{1}^{'}$ the nonlinear coefficient of the cavity 1, and $\gamma_{2}$ is the nonlinear coefficient of the cavity 2; we have defined $\kappa_{\rm c}={\rm arcsin}(\sqrt{\theta_{\rm c}})/L_2$ with $\theta_c$ the power coupling ratio between the two ring cavities, and $\kappa_{1}=\sqrt{\theta_{1}}/L_2$.

Usually, once a device are fabricated, the loss of the system $\alpha_{1,2}$, the group velocity mismatch $\Delta k^{'}$, the group velocity dispersion $k_{1,2}^{''}$, and the nonlinear coefficient $\gamma_{1,2}$ as well as the power coupling ratio $\kappa_1$ and $\kappa_{\rm c}$ are not tunable, unless the device is assisted by special materials or technologies. For example, by coupling graphene to a nitride photonic microresonator, the dispersion of the device can be tuned due to the complex optical conductivity of the graphene, which can be tuned through an external electric field \cite{Yao2018}. The parameter $A_{\rm in}$ corresponding to the pump field can be tuned by changing the external pump power, such a scheme has been used for overcoming the thermal optical effect to stabilize the generated Kerr OFCs \cite{Brasch2016a}. Especially, in the case of Kerr OFC generation in a single cavity, the frequency detuning are usually tuned across a cavity resonance to first generate OFCs and finally enter the soliton state \cite{Herr2014}. The frequency detuning is proportional to the difference between the resonant frequency and the pump frequency, and inversely proportional to the free spectral range ({\rm FSR}), i.e., $\delta_{1,2}=(\omega_{\rm r1,2}-\omega_{\rm p})/({\rm FSR}_{1,2} L_2)$ with $\omega_{\rm r1,2}$ the resonant frequencies closest to the pump frequency $\omega_{\rm p}$. By controlling the temperature of the cavity 1 and the cavity 2 through thermo-electric heaters (or coolers) resulting in shifting of the resonant frequencies $\omega_{\rm r1,2}$, the detunings of the cavity 1 and the cavity 2 can be independently tuned. Indeed, such a scheme has been used in previous works \cite{Miller2015, Lu2019}. In addition to shifting the resonant frequencies, sweeping the pump frequency \cite{Herr2014, Guo2017} can tune the detunings of the cavity 1 and the cavity 2 simultaneously. In our numerical simulation in Sec.~\ref{sec:level4}, the detunings of the cavity 1 and the cavity 2 are tuned simultaneously and independently in Fig.~\ref{fig:5} and in Fig.~\ref{fig:6}, respectively.

With the normalization such that $z\rightarrow z\alpha_{2}$, $\tau\rightarrow\tau\sqrt{2\alpha_{2}/|k_{2}^{''}|}$, $F_{1,2}=A_{1,2}\sqrt{\gamma_2/\alpha_{2}}$, $d=\Delta k^{'}\sqrt{2/|k_{2}^{''}|\alpha_{2}}$, $\alpha=\alpha_{1}/\alpha_{2}$, $\Delta_{1,2}=\delta_{1,2}/\alpha_{2}$, $\eta_{1}=k_{1}^{''}/|k_{2}^{''}|$, $\eta_{2}={\rm sgn}(k_{2}^{''})$, $\Gamma=\gamma_{1}/\gamma_{2}$, $\kappa=\kappa_{\rm c}/\alpha_{2}$, and $S=\kappa_{1}\sqrt{\gamma_{2}/\alpha_{2}^3}A_{\rm in}$, Eqs.~(\ref{eq:1}) and (\ref{eq:2}) can be rewritten in the normalized form
\begin{eqnarray}
\frac{\partial F_{1}}{\partial z}=&&\left[-\alpha-i\Delta_{1}-d\frac{\partial}{\partial \tau}-i\eta_{1}\frac{\partial^2}{\partial\tau^2}+i\Gamma|F_{1}|^2\right]F_{1}\nonumber\\
&&+i\kappa F_{2}+iS,
\label{eq:3}
\end{eqnarray}
\begin{equation}
\frac{\partial F_{2}}{\partial z}=\left[-1-i\Delta_{2}-i\eta_{2}\frac{\partial^2}{\partial\tau^2}+i|F_{2}|^2\right]F_{2}+i\kappa F_{1},
\label{eq:4}
\end{equation}
Note that the normalization implies that the cavity 2 must be nonlinear, i.e., $\gamma_2\neq0$. The equations can be numerically integrated over the propagation distance $z$ with a split-step Fourier method \cite{Sinkin2003}. Because Eqs.~(\ref{eq:3}) and (\ref{eq:4}) (as well as Eqs.~(\ref{eq:1}) and (\ref{eq:2})) are derived in a reference frame that moves at the group velocity of the light in the cavity 2, the temporal window of the Fourier transforms in simulations is the (normalized) roundtrip time of the cavity 2, and hence the frequency grid corresponds to the FSR of the cavity 2.

\begin{figure}[tb]
\includegraphics[width=240pt]{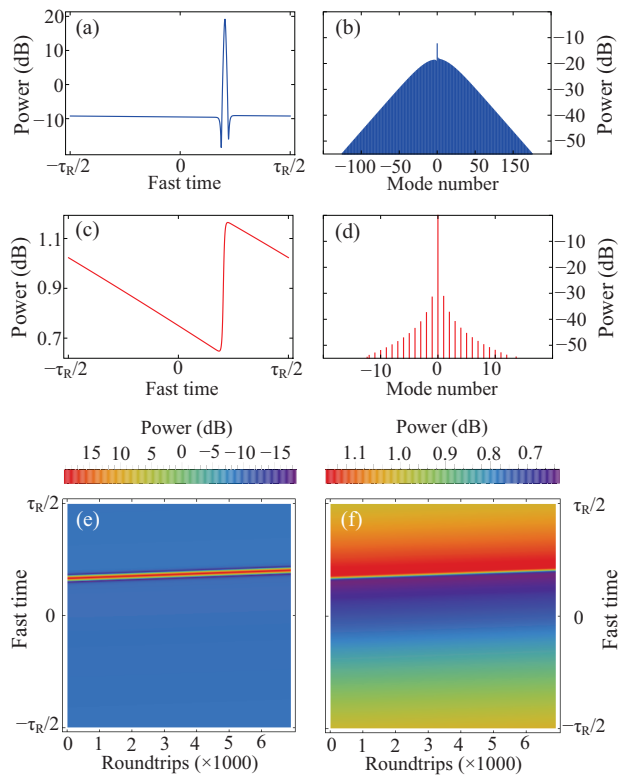}
\caption{\label{fig:2}Intracavity temporal profiles and corresponding comb spectra. (a, c) The DKS in the cavity 2 and the temporal profile in the cavity 1. (b, d) Corresponding comb spectra. (e) and (f) The evolution of the DKS in the cavity 2 and the temporal profile in the cavity 1, respectively.}
\end{figure}
To validate Eqs.~(\ref{eq:3}) and (\ref{eq:4}) we show a simulated DKS in Fig.~\ref{fig:2}. The simulation is performed with followed parameters similar to the ones used in Ref.~\cite{Xue2019}, but we have ideally assumed that the cavity 1 is linear: $d=-328.07$, $\alpha=1.17$, $\Delta_{2}=39.33$, $\Delta_{1}=2.88$, $\eta_{2}=-1$, $\eta_{1}=1$, $\Gamma=0$, $\kappa=12.05$, and $S=2.04$. The normalized roundtrip time of the cavity 2 is set by $\tau_{\rm R}=1/{\rm FSR_2}=39.08$. The cavity 1 has normal dispersion, while the cavity 2 has anomalous dispersion and is nonlinear. From the functional point of view, the cavity 1 is only used for the pumping recycle due to its normal dispersion and the zero nonlinearity, and the cavity 2 is used to provide Kerr nonlinearity and sustain the DKS due to its anomalous dispersion. To excite the DKS, a shot gaussian pulse $F_{\rm gaussian}=\sqrt{P_0}\exp(-\tau^2/2\tau_0^2)$ with normalized half width $\tau_0=1.56$ and peak power $P_0=98.23$ is introduced into the cavity 2 after the system reaches the cw steady state. The gaussian pulse quickly evolves into a DKS on a cw background (Fig.~\ref{fig:2}(a)), and the spectrum is correspondingly rapidly expanded (Fig.~\ref{fig:2}(b)). The soliton and its spectrum is slightly unsymmetrical even though we only consider the second-order dispersion. We attribute it to the nonzero group velocity mismatch $d$ between the cavity 1 and the cavity 2. For the optical field in the cavity 1, the temporal profile behaves a small step-like shape (Fig.~\ref{fig:2}(c)), and the corresponding spectrum only contains few comb lines and has a concave envelope (Fig.~\ref{fig:2}(d)), which is similar to the one obtained from an orthogonally polarized dual-pumped single cavity via the cross-phase modulation (XPM) effect \cite{Bao2019}. Since the cavity 1 is linear, the sideband comb lines in the cavity 1 are obtained from the cavity 2 through the coupling of the two cavities rather than from the self-phase modulation (SPM) effect. The step-like temporal profile is similar to the second harmonic temporal pattern found in quadratic OFCs generation in a doubly resonant cavity \cite{Leo2016}. We find that although the DKS is stable, it has a temporal drift that shares a same speed with the temporal profile of the optical field in the cavity 1, and the DKS and the ``step" are in the same position of the temporal window (Fig.~\ref{fig:2}(e) and (f)). This result implies the group velocity of the DKS is changed by the interaction of the optical field in the two cavities. Although the XPM effect is absent here, it is similar to a phenomenon termed ``soliton trapping" having been studied in a microring resonator \cite{Suzuki2019}, which describes two optical pulses with different initial group velocities will finally propagate with same group velocity through the XPM effect.

\section{\label{sec:level3}Modulation instability analysis}
Similar to single cavity systems, MI of the cw steady state plays an important role for frequency-comb formation in dual-coupled cavities systems. When the dual-coupled cavities system is coherently pumped, assuming the modulation instability has not occurred, the system will soon stay at the cw steady state. By setting all the derivatives to zero in Eqs.~(\ref{eq:3}) and (\ref{eq:4}), one would obtain the cw steady state intracavity power satisfying the following coupled equations
\begin{equation}
\kappa^2 Y_1=Y_2^3-2\Delta_2Y_2^2+(\Delta_2^2+1)Y_2,
\label{eq:5}
\end{equation}
\begin{equation}
\kappa^2X=c_3 Y_2^3+c_2 Y_2^2+c_1 Y_2,
\label{eq:6}
\end{equation}
where $Y_1=|F_{1\rm{s}}|^2$, $Y_2=|F_{2\rm{s}}|^2$, and $X=|S|^2$ correspond to the optical power in the cavity 1, the optical power in the cavity 2 and the pump power, respectively, with $F_{1\rm{s}}$ and $F_{2\rm{s}}$ the cw stationary solutions. The polynomial coefficients in Eq.~(\ref{eq:6}) are defined as $c_1=\left(1+\Delta_2^2\right)\bar{Y}_1^2+2\kappa^2\Delta_2\bar{Y}_1+\left(\alpha+\kappa^2\right)^2+\alpha^2\Delta_2^2$, $c_2=-2\left(\Delta_2\bar{Y}_1^2+\kappa^2\bar{Y}_1+\alpha^2\Delta_2\right)$, and  $c_3=\bar{Y}_1^2+\alpha^2$, with $\bar{Y}_1=\Gamma Y_1-\Delta_1$. By treating the optical field in the cavity 2 as a pump of the cavity 1, Eq.~(\ref{eq:6}) is identical with the well-known cubic equation for a single Kerr cavity \cite{Haelterman1992}. The cubic equation is single valued for the normalized detuning $\Delta_2\leq\sqrt{3}$, while it may has one, two, or three values and has a bistable hysteresis ``S" shape for the normalized detuning $\Delta_2>\sqrt{3}$ \cite{Godey2014}. The middle value is always a forbidden value because it is unstable with respect to cw perturbations. However, for the case of dual-coupled cavities, solutions of Eq.~(\ref{eq:5}) and Eq.~(\ref{eq:6}) are more complex because they admit at most 9 different values for they are equivalent to a 9th degree algebraic equation with respect to $Y_2$. Although the MI for a single cavity system has been well studied \cite{Hansson2013, Godey2014}, it is improper to simply treat the optical field in the cavity 1 as a pump of the cavity 2 due to the mode coupling and the group velocity mismatch  between the two cavities.

\begin{figure}[tb]
\includegraphics[width=240pt]{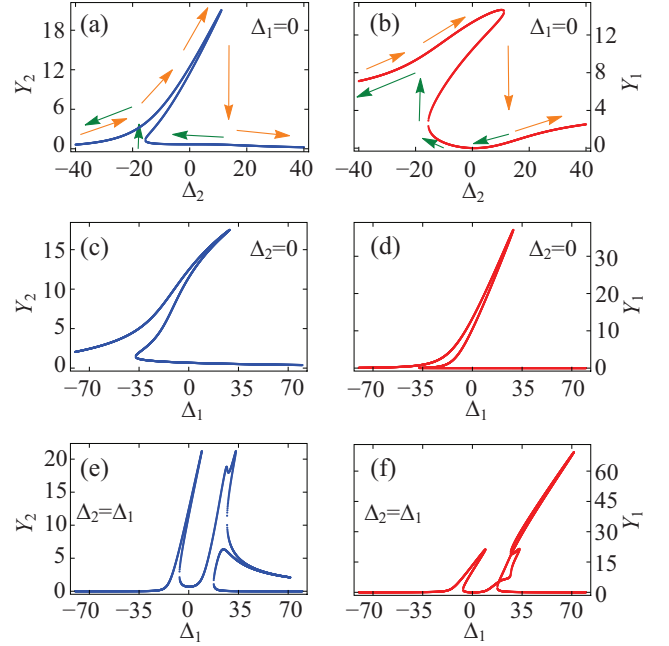}
\caption{\label{fig:3} Continuous-wave steady state intracavity power as a function of detunings. The parameters $\alpha=1.17$, $\Gamma=1$, $\kappa=12.05$, and $S=10$ are used for computing. (a) $Y_1$ and (b) $Y_2$ versus $\Delta_2$, with $\Delta_1=0$. The orange (green) arrows show the intracavity power path assuming MI does not occur, when $\Delta_2$ is tuned from blue (red) detuning to red (blue) detuning. (c) $Y_1$ and (d) $Y_2$ versus $\Delta_1$, with $\Delta_2=0$. (e) $Y_1$ and (f) $Y_2$ versus $\Delta_1$, with $\Delta_2=\Delta_1$. The intracavity power shows tilted curves caused by the Kerr nonlinearity.}
\end{figure}

Fig.~\ref{fig:3} shows the intracavity power for different detunings. In Fig.~\ref{fig:3}(a) and (b), both $Y_1$ and $Y_2$ exhibit a nonlinear tilt of the resonance with the detuning of the cavity 1 fixed to zero. $Y_1$ is intuitively small when the cavity 2 has a large red detuning, while $Y_1$ remains a relatively large value when the cavity 2 has a large blue detuning. For the situation that the detuning of the cavity 2 is fixed to zero, the intracavity power $Y_1$ and $Y_2$ also show a tilted resonance (Fig.~\ref{fig:3}(c) and (d)). For $\Delta_2=\Delta_1$, which can be achieved by the use of scanning the pump frequency assuming that the cavity 1 is identical to the cavity 2, the cavity resonance splits into two resonances both with a tilted shape, and the red shifted resonance is extremely distorting due to the interaction of the two cavities (Fig.~\ref{fig:3}(e) and (f)). It is identical to that two coupled linear cavities have a ``W" shape transmission spectrum indicating that there are two resonances around the center frequency caused by the mode coupling between the two cavities.

The MI is investigated by using the ansatz $F_1=F_{1\rm{s}}+a$ and $F_2=F_{2\rm{s}}+b$ for the perturbation of cw stationary solutions, with $a=a_1\exp{\left(\lambda z+i\Omega \tau\right)}+a_2\exp{\left(\lambda^* z-i\Omega \tau\right)}$, $b=b_1\exp{\left(\lambda z+i\Omega \tau\right)}+b_2\exp{\left(\lambda^* z-i\Omega \tau\right)}$ and $\Omega$ the modulation frequency. After introducing the ansatz into Eq.~\ref{eq:3} and Eq.~\ref{eq:4} and linearizing with respect to $a$ and $b$, we obtain a $4\times4$ matrix equation in the Fourier domain whose eigenvalues satisfy the following equation
\begin{equation}
\lambda^4+C_3\lambda^3+C_2\lambda^2+C_1\lambda+C_0=0,
\label{eq:7}
\end{equation}
where the polynomial coefficients are defined as
\begin{eqnarray}
C_0=&&\kappa^2\left(\kappa^2+2\bar{\alpha}-2\bar{\Delta}_1\bar{\Delta}_2\right)\nonumber\\
&&+\left(\bar{\alpha}^2+\bar{\Delta}_1^2-\Gamma^2Y_1^2\right)\left(1-Y_2^2+\bar{\Delta}_2^2\right)\nonumber\\
&&+2\Gamma Y_2\left[1-\left(\Delta_2-Y_2\right)^2\right],\nonumber
\end{eqnarray}
\begin{eqnarray}
C_1=2\left[\bar{\alpha}\left(1-Y_2^2+\kappa^2+\bar{\alpha}+\bar{\Delta}_2^2\right)
+\bar{\Delta}_1^2+\kappa^2-\Gamma^2Y_1^2\right],\nonumber
\end{eqnarray}
\begin{eqnarray}
C_2=1-Y_2^2-\Gamma^2Y_1^2+2\kappa^2+\bar{\alpha}\left(4+\bar{\alpha}\right)+\bar{\Delta}_1^2+\bar{\Delta}_2^2,\nonumber
\end{eqnarray}
\begin{eqnarray}
 C_3=2\left(1+\bar{\alpha}\right),
 \end{eqnarray}
where $\bar{\Delta}_1=\Delta_1-\eta_1 \Omega^2-2\Gamma Y_1$, $\bar{\Delta}_2=\Delta_2-\eta_2 \Omega^2-2Y_2$, and $\bar{\alpha}=\alpha-id\Omega$.

\begin{figure}[tb]
\includegraphics[width=240pt]{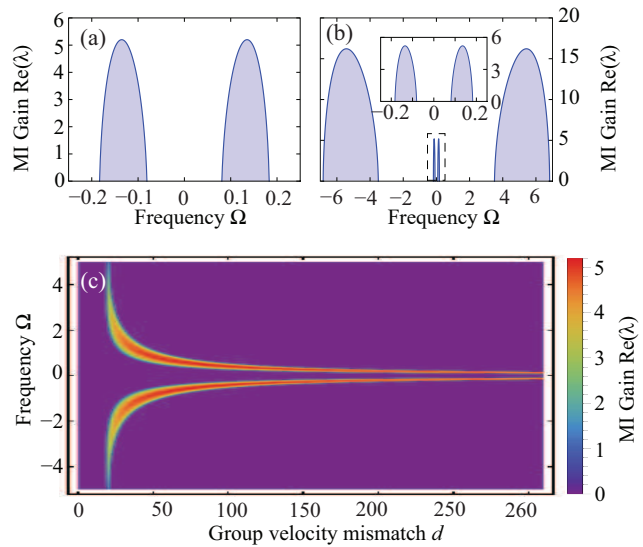}
\caption{\label{fig:4}MI gains calculated with parameters as follow: $d=-300$, $\Delta_1=5$, and $\Delta_2=5$, and other parameters are the same as in Fig~\ref{fig:3}. Note that under these parameters the cw steady state is multi-stable, and the modulation instability gains are computed on the upper branch. (a) $\eta_1=1$ and $\eta_2=1$, i.e., both the cavity 1 and the cavity 2 have normal dispersion. (b) $\eta_1=1$ but $\eta_2=-1$, i.e., the cavity 1 has normal dispersion but the cavity 2 has anomalous dispersion. The inset in (b) is zoom-in MI gain identical to the one in (a). (c) MI gain spectrum for a wide range of group velocity mismatch $d$, with dispersion parameters the same as in (b).}
\end{figure}

The MI gain, i.e., $\rm{Re}{(\lambda)}$, depends not only on the parameters of the pump and the cavities, but also on which branch of the cw steady state the system is in, which is determined by the detuning scanning direction. For example, as the orange arrows showed in Fig.~\ref{fig:3}(a) and (b), if $\Delta_2$ is tuned from blue (negative) detuning to red (positive) detuning, the intracavity power will first stay at the upper branch. Until the inflection point is crossed, it drops down to the lower branch. On the contrary, if $\Delta_2$ is tuned from red detuning to blue detuning, the intracavity power will first pass through the lower branch and then jump to the upper branch (green arrows in Fig.~\ref{fig:3}(a) and (b)). Fig.~\ref{fig:4} shows the MI gain calculated on the upper branch. In Fig.~\ref{fig:4}(a), both the cavity 1 and the cavity 2 have normal dispersion, while in Fig.~\ref{fig:4}(b), the cavity 1 has normal dispersion but the cavity 2 has anomalous dispersion. It has been known that for single cavity systems, normal dispersion will suppress MI and thus is detrimental to the formation of OFCs \cite{Chembo2010, Hansson2013}. However, it have been experimentally demonstrated that two coupled normal dispersion cavities are also able to generate OFCs due to the local anomalous dispersion caused by mode coupling effect \cite{Xue2015a}. Our MI analysis indeed shows that it is possible for comb generation for two coupled cavities both having normal dispersion (see Fig.~\ref{fig:4}(a)). Here, for parameters used in Fig.~\ref{fig:4}, we attribute the MI gain to the nonzero group velocity mismatch $d$, a parameter that is absent in LLE \cite{Coen2013}. Fig.~\ref{fig:4}(c) shows the MI gain spectrum calculated by a wide range of group velocity mismatch $d$, and other parameters are same as in Fig.~\ref{fig:4}(a). When $d$ is very small, MI does not occur. However, as $d$ increases, MI occurs at frequencies far away from the pump frequency. As $d$ continues to increase, the MI gain bandwidth is both shifted inward and decreased. Until the bandwidth becomes so narrow that cannot cover at least one resonance, the MI gain eventually disappears. The group velocity mismatch induced MI also occurs in quadratical OFCs \cite{Leo2016, Leo2016a, Hansson2017}. We note that although we highlight the importance of the group velocity mismatch in the case that both the two cavities have the normal dispersion, it is still possible for MI occuring with $d=0$ and certain detunings, for example, $d=0$, $\Delta_1=7$ and $\Delta_2=30$ with other parameters same as in Fig.~\ref{fig:4}(a). Compared with Fig.~\ref{fig:4}(a), the MI gain shown in Fig.~\ref{fig:4}(b) has an additional broader sideband caused by the anomalous dispersion of the cavity 2, with a narrow sideband near the pump frequency exactly the same as in Fig.~\ref{fig:4}(a). Clearly, the group velocity mismatch $d$ influences the MI at a same extent even though the anomalous dispersion is present.

\section{\label{sec:level4}Comb spectrum simulation}
Unlike single cavity systems, in which only the anomalous dispersion is friendly to the formation of OFCs \cite{Chembo2010, Hansson2013}, our MI analysis in the last section indicates dual-coupled cavities systems with arbitrary dispersion have the potential to promote MI leading to sideband growth, then resulting in the generation of OFCs with the help of cascaded four-wave mixing (FWM). We are most interest in two cases that are ``$+/+$" (both the cavity 1 and the cavity 2 have normal dispersion) and ``$+/-$" (the cavity 1 has normal dispersion while the cavity 2 has anomalous dispersion).

\begin{figure*}[tb]
\includegraphics[width=400pt]{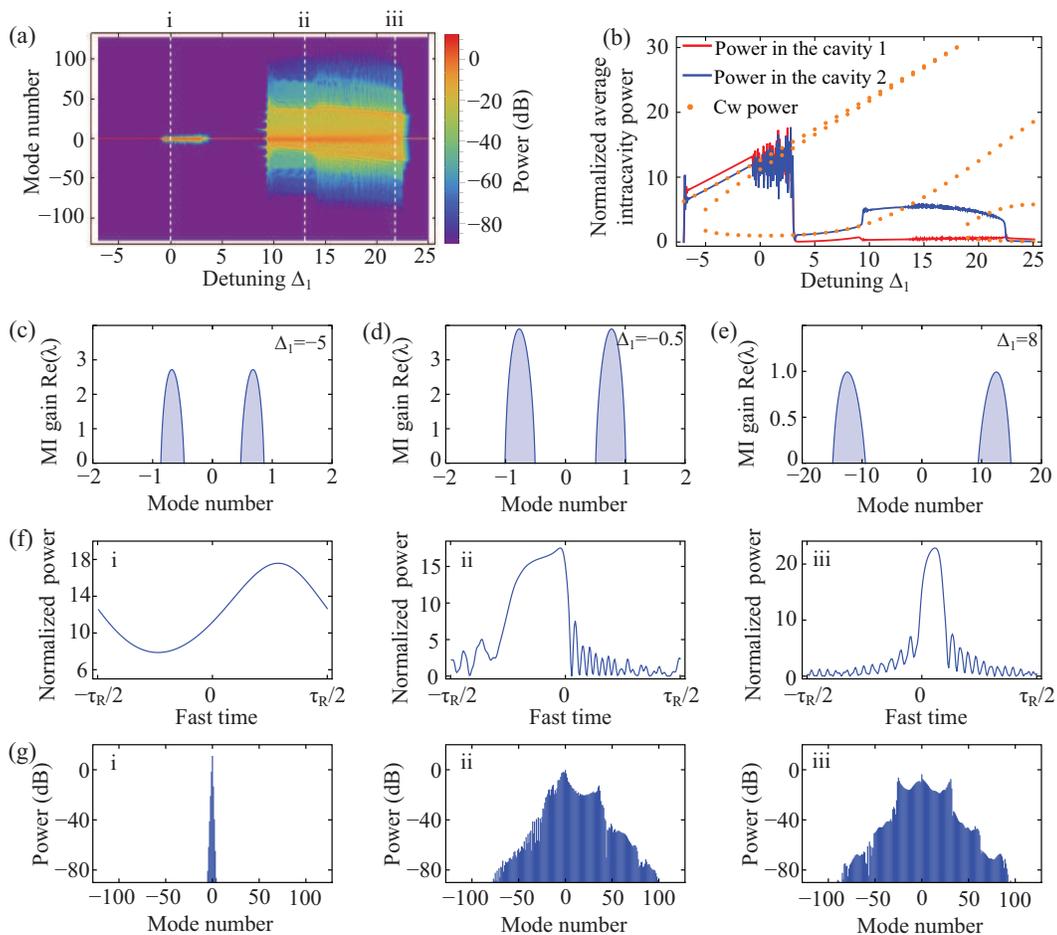}
\caption{\label{fig:5}The simulated comb for the ``$+/+$" case. (a) Evolution of the comb spectrum in the cavity 2 with the normalized detuning $\Delta_1$. (b) Normalized average intracavity power in the cavity 1 (red line) and in the cavity 2 (blue line). Orange dots denote the cw steady power in the cavity 2 calculated by Eqs.~(\ref{eq:5}) and (\ref{eq:6}). (c-e) The MI gain at $\Delta_1=-5$, $\Delta_1=-0.5$ and $\Delta_1=8$, respectively. (f) and (g) Temporal profiles and corresponding comb spectra in the cavity 2 for the different positions {\romannumeral1}-{\romannumeral3} denoted by white lines in (a).}
\end{figure*}
\begin{figure*}[tb]
\includegraphics[width=400pt]{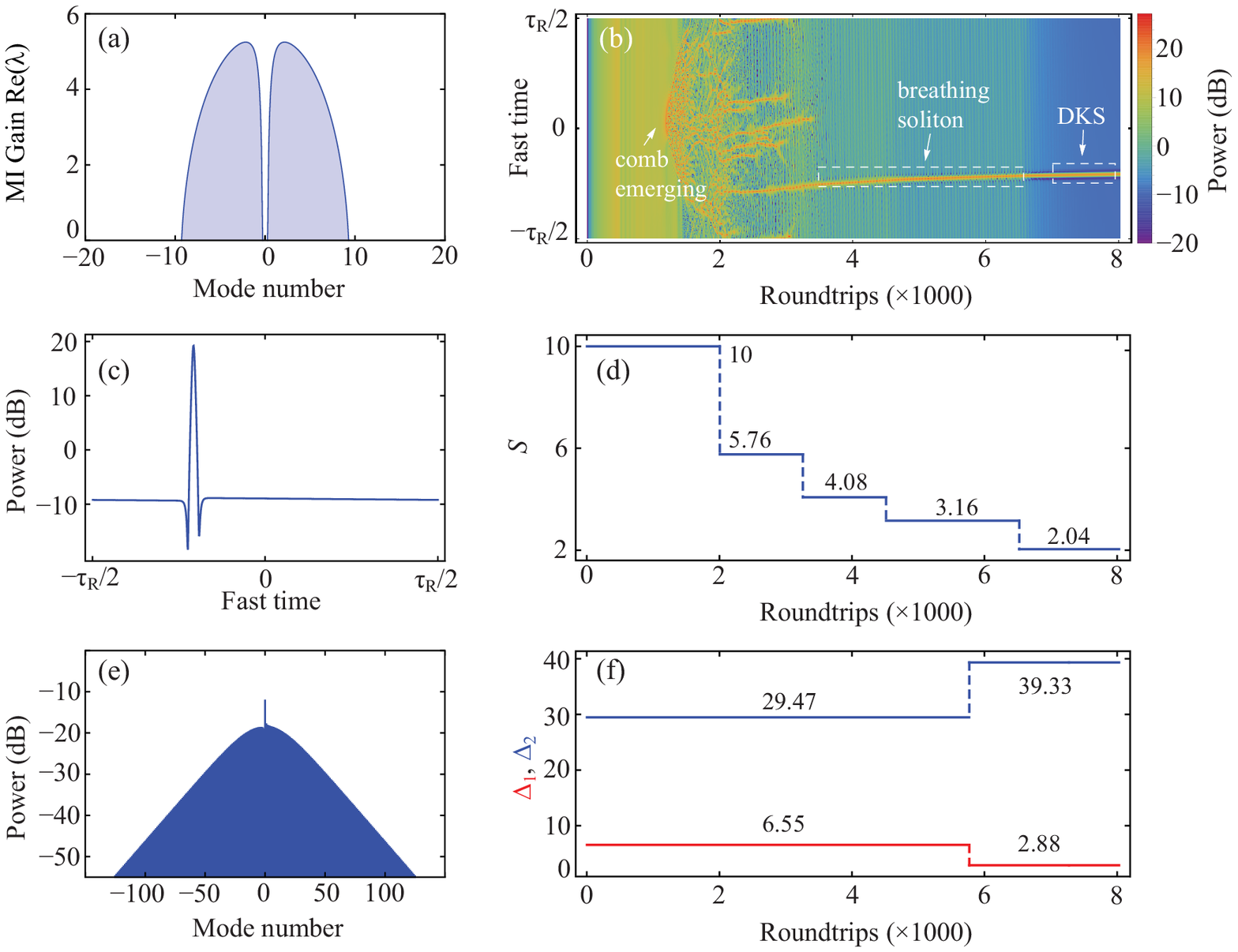}
\caption{\label{fig:6}The simulated comb for the ``$+/-$" case. (a) The MI gain. (b) Evolution of the temporal profile in the cavity 2 with roundtrips. (c) and (e) A single DKS and the corresponding comb spectrum. (d) and (f) Pump modulation and detuning tuning scheme.}
\end{figure*}
As an example for the ``$+/+$" case, we perform a simulation with a detuning scanning scheme. The simulation is started from $\Delta_1=-7$, and then the detuning $\Delta_1$ is slowly increased. The normalized roundtrip time of the cavity 2 is set by $\tau_{\rm R}=39.1$ corresponding to the normalized free spectral range ${\rm FSR}_2=1/\tau_{\rm R}\approx 0.026$. In the simulation, we assume ${\rm FSR}_1={\rm FSR}_2$ and $\Delta_2=\Delta_1$ throughout. The dispersion of the two cavities is normal and has a same magnitude, i.e., $\eta_1=\eta_2=1$, and the group velocity mismatch is, however, nonzero and large, $d=-300$. The coupling strength between the cavity 1 and the cavity 2 and the propagation losses ratio of the two cavities are $\kappa=12.05$ and $\alpha=1.17$, respectively. The remained parameters are set by $\Gamma=1$ and $S=12$. Fig.~\ref{fig:5}(a) shows the evolution of the comb spectrum in the cavity 2 with the normalized detuning $\Delta_1$. Fig.~\ref{fig:5}(b) shows the average intracavity power of the cavity 1 (red line), the average intracavity power of the cavity 2 (blue line) and cw steady state power of the cavity 2 (orange dot) calculated by Eqs.~(\ref{eq:5}) and (\ref{eq:6}). As the normalized detuning $\Delta_1$ increased, in the beginning, the system enter into upper branch of the cw steady state. We plot the MI gain at $\Delta_1=-5$ as shown in Fig.~\ref{fig:5}(c). Although the MI gain is nonzero with modulation frequencies, the system is still in the cw state because the allowed mode number must be an integer. As $\Delta_1$ increased still further, the MI gain increases and shifts far from the pump mode. Around $\Delta_1=-0.5$, the MI gain band covers the first mode (see Fig.~\ref{fig:5}(d)), and the comb sideband at the first mode begins growing. Subsequently, a narrow OFC is generated through the cascaded FWM (Fig.~\ref{fig:5}(f and g, \romannumeral1)). Although the comb is very narrow, we find that the corresponding average intracavity power is rapidly and randomly oscillating, which is similar to the chaos state in the single cavity configuration \cite{Godey2014}. Around $\Delta_1\approx2.5$ the narrow comb vanishes and the system drops to the lower cw branch. The MI gain at $\Delta_1=8$ (Fig.~\ref{fig:5}(e)) covers several modes, leading to primary combs growing around these modes, and after that the spectrum rapidly expands due to the cascaded FWM, with comb lines spacing by a single FSR. The phenomenon of discontinuous evolution of the comb spectrum also occurs in the simulations in whispering-gallery-mode microresonators with backscattering \cite{Kondratiev2020}. We find that the average intracavity power of the cavity 2 increases with the generation of the broadband comb, while the average intracavity power of the cavity 1 decreases leading to extremely low power remained in the cavity even though it is coupled with the pump. Fig.~\ref{fig:5}(f and g, \romannumeral2) show an OFC at $\Delta_1=12$. The envelope of the comb spectrum is asymmetric. Comb lines with power larger than $ -20$ $\rm{dB}$ in the right side are more than the ones in the left side. Around $\Delta_1\approx14$ the comb transitions to an obviously unstable regime, where comb lines and the average inracavity power rapidly oscillate. As $\Delta_1$ increased further, the system accesses a narrow stable regime before the regime where the comb vanishes and drops to the cw steady state. Fig.~\ref{fig:5}(f and g, \romannumeral3) show an OFC at $\Delta_1=21.75$. The corresponding temporal profile in the cavity 2 seems like a solitonic pulse ``platicon" \cite{Lobanov2015}, which can be generated in a single cavity resonator in the normal dispersion regime with a local dispersion perturbation.

In the simulation for the ``$+/-$" case, we assume the parameter $S$ and detunings $\Delta_1$ and $\Delta_2$ can be flexibly and independently tuned with other parameters same as in Fig.~\ref{fig:2}. To stimulate the primary comb through MI, the simulation is started with $S=10$, $\Delta_1=6.55$ and $\Delta_2=29.47$, and the corresponding MI gain is shown in Fig.~\ref{fig:6}(a). Compared with the ``+/+" case using the detuning scanning scheme, the comb for the ``$+/-$" case is rapidly expanded due to the anomalous dispersion provided by the cavity 2, and stays at the chaos state until a soliton generated, leading to a featureless comb spectrum evolution. However, the evolution of the temporal profile in the cavity 2 shown in Fig.~\ref{fig:6}(b) reveals a distinct dynamics. After the system obtains power from the pump, the primary comb lines emerge around the first mode predicted by the MI gain, leading to an initial waveform with a single peak. In contrast, in a single cavity with anomalous dispersion the MI gain is usually far from the pump mode resulting in an initial waveform with periodic multi peaks \cite{Chembo2010, Godey2014, Herr2014}. With the comb spectrum expanded through the cascaded FWM, the initial single peak is broken up and evolves into several peaks. We gradually decrease the value of the paramater $S$ (corresponding to decreasing the pump power) from $S=10$ to $S=2.04$ as shown in Fig.~\ref{fig:6}(d). As $S=4.08$, one of the peaks evolves into a breathing soliton \cite{Godey2014, Peng2019} while the others collapse into the cw background. To stabilize the soliton, we finally set $\Delta_1=2.88$ and $\Delta_2=39.33$. The generated stable single soliton and the corresponding spectrum are shown in Fig.~\ref{fig:6}(c) and (d), respectively. The DKS generated by the pump modulation and detuning tuning scheme is identical to the one excited by a high peak power gaussian pluse in Fig.~\ref{fig:2}.

\section{\label{sec:level5}Conclusion}
In conclusion, we have introduced two normalized coupled mean-field equations describing the optical field in dual-coupled cavities systems, which are validated by a single DKS in the cavity 2 excited by introducing a gaussian pulse into the cavity 2. By setting all derivatives in the equations to zero, we obtain the solutions of cw steady state, which show two tilde resonances induced by the mode coupling between the two cavities and Kerr nonlinearity. We also investigated MI of the cw steady state, and the group velocity mismatch is found play an important role for MI. We performed a numerical simulation for the case that both the two cavities have normal dispersion. With detunings of the two cavities simultaneously scanned from blue detuning to red detuning, a platicon generated in the cavity 2. We also performed a numerical simulation for the case that the cavity 1 has normal dispersion while the cavity 2 has anomalous dispersion. With the pump power and detunings of the two cavities tuned, a single DKS is generated in the cavity 2. MI of the cw steady state in cavities is key to understanding the Kerr OFC generation. Recent researches focusing on dual-coupled cavities systems have revealed novel Kerr OFC dynamics \cite{Helgason2021, Tikan2021}. This work theoretically demonstrates the possibility for Kerr OFC generation beginning with MI in dual-coupled cavities systems.

\begin{acknowledgments}
This work was funded by the State Key Laboratory of Quantum Optics and Quantum Optics Devices, Shanxi University, Shanxi, China (KF202004).
\end{acknowledgments}

\bibliographystyle{apsrev4-2}
\bibliography{bib}

\end{document}